\documentclass[]{JHEP3}
 \pdfoutput=1
\usepackage{epsfig}
\def\be{\begin{equation}}
\def\ee{\end{equation}}
\def\bea{\begin{array}}
\def\eea{\end{array}}
\def\beqa{\begin{eqnarray}}
\def\eeqa{\end{eqnarray}}
\def\beqas{\begin{eqnarray*}}
\def\eeqas{\end{eqnarray*}}

\def\bp{\begin{picture}}
\def\ep{\end{picture}}
\def\bc{\begin{center}}
\def\ec{\end{center}}
\def\bfig{\begin{figure}}
\def\efig{\end{figure}}

\def\bit{\begin{itemize}}
\def\eit{\end{itemize}}
\def\nn{\nonumber}
\def\f{\frac}

\def\[{\left[}
\def\]{\right]}
\def\({\left(}
\def\){\right)}

\def\..{\left.}
\def\.{\right.}
\def\tl{\tilde}
\def\ra{\rightarrow}
\def\la{\leftarrow}

\def\tm{\times}

\def\da{\dagger}

\def\la{\lambda}

\def\al{\alpha}

\def\ka{\kappa}

\def\ep{\epsilon}

\def\pa{\partial}
\def\pr{\prime}

\title{ NMSSM From Alternative Deflection in Generalized Deflected Anomaly Mediated SUSY Breaking}
\author{ Xiaokang Du$^1$, Fei Wang$^{1,2}$\\
$^1$ School of Physics, Zhengzhou University, 450000,ZhengZhou
P.R.China\\
$^2$ State Key Laboratory of Theoretical Physics, Institute of Theoretical Physics,
                Chinese Academy of Sciences, Beijing 100080, P. R. China
}

\abstract{ We propose a new approach to generate messenger-matter interactions in deflected anomaly mediated SUSY breaking mechanism from typical holomorphic messenger-matter mixing terms in the Kahler potential. This approach is a unique feature of AMSB and has no analog in GMSB-type scenarios. New coupling strengths from the scaling of the (already known) Yukawa couplings always appear in this approach. With messenger-matter interactions in deflected AMSB, we can generate a realistic soft SUSY breaking spectrum for next-to-minimal supersymmetric standard model(NMSSM). Successful electroweak symmetry breaking conditions, which is not easy to satisfy in NMSSM for ordinary AMSB-type scenario, can be satisfied in a large portion of parameter space in our scenarios. We study the relevant phenomenology for scenarios with (Bino-like) neutralino and axino LSP, respectively.
 In the case of axino LSP, the SUSY contributions to $\Delta a_\mu$ can possibly account for the muon $g-2$ discrepancy. The corresponding gluino masses, which are found to below 2.2 TeV, could be tested soon at LHC.
}
\begin{document}
\maketitle \indent
\newpage
\section{Introduction}
  Weak scale supersymmetry(SUSY) is one of the most interesting candidates for new physics beyond the standard model(SM).
  Its low energy phenomenology is determined mainly by the relevant soft SUSY breaking parameters
  which are required to preserve flavor and CP with a good accuracy.
  Such soft SUSY breaking parameters can be predicted by the mediation mechanism of SUSY breaking.
  So it is crucial to understand various well-motivated mediation mechanisms, for example,
 the gauge mediated SUSY breaking(GMSB)\cite{GMSB}, the anomaly mediated SUSY breaking(AMSB)\cite{AMSB} mechanisms.

  Minimal AMSB, which is determined solely by the parameter $F_\phi\simeq m_{3/2}$,
is insensitive to the UV theory\cite{AMSB:RGE} and predicts a flavor conservation soft SUSY breaking spectrum.
Unfortunately, negative slepton squared masses will appear and the minimal scenario must be extended.
Although there are many possible ways to tackle such tachyonic slepton problem, the most elegant
solution from aesthetical point of view is the deflected AMSB\cite{AMSB:deflect,Nelson:2002sa}(dAMSB) scenario.
In deflected AMSB, additional messenger sectors are introduced to deflect the AMSB trajectory and
additional gauge mediation contributions can possibly push the negative slepton squared masses to positive values\cite{dAMSB:example}.
On the other hand, $N\geq 4$ species are always needed to give positive slepton squared masses with naturally negative deflection parameters,
possibly leading to strong gauge couplings below GUT scale or Landau pole below Planck scale. Besides, electroweak naturalness and the discovered 125 GeV Higgs boson by both the ATLAS\cite{ATLAS:higgs} and CMS collaborations\cite{CMS:higgs} of LHC,
may indicate respectively the lightness of stop and large trilinear coupling $A_t$;
null search results of sparticles  with 36 $fb^{-1}$ of data at the (13 TeV) LHC by the ATLAS and CMS collaborations
\cite{36fb:colored,36fb:EW} suggest that the low energy SUSY spectrum should display an intricate pattern, for example, the first two generation squarks need to be heavy to avoid the stringent constraints from LHC. However, spectrum of such a type will in general not be predicted in ordinary (d)AMSB scenarios.

We had proposed to introduce general messenger-matter interactions in dAMSB to solve the previous problems\cite{Fei:1508.01299} which can be advantageous in various aspects.
The previous formalism in \cite{Fei:1508.01299,Fei:1602.01699}, within which the messenger-matter interactions are introduced in the superpotential,
has an analog in GMSB\cite{GMSB:MM}. In this paper, we propose an alternative approach to including messenger-matter interactions in (d)AMSB
from typical holomorphic terms in the Kahler potential. Such an approach, with the messenger scales set by the compensator vacuum expectation value(VEV),
can only be possible in AMSB type scenarios.

 Next-to-Minimal Supersymmetric Standard Model(NMSSM)\cite{NMSSM} is a singlet extension of MSSM that has various advantages.
 The SUSY preserving $\mu$ parameter, which (however) needs to lie near the soft SUSY breaking scale to trigger EWSB,
 could be generated with the correct scale once the singlet scalar acquires a VEV.
Besides, with additional tree-level contributions or through doublet-singlet mixing,
NMSSM can easily accommodate the discovered 125 GeV Higgs boson without much fine-tuning.
Low energy NMSSM from typical SUSY breaking mechanism, such as GMSB,
is always bothered by the requirement to achieve successful EWSB with suppressed trilinear couplings $A_\ka,A_\la$ and $m_S^2$,
rendering the model building non-trivial\cite{NMSSM:GMSB}.
Such difficulty always persists in ordinary (d)AMSB-type scenarios.
We find that phenomenological interesting NMSSM spectrum can be successfully generated
 from non-trivial holomorphic messenger-matter mixing terms in the Kahler potential.
Besides, the discrepancy between the theoretical predictions for the muon anomalous
magnetic momentum and the experiments, can possibly be solved in our scenario.

This paper is organized as follows.
In Sec \ref{secII}, we propose our new scenario and discuss the general methods to obtain the soft SUSY parameters.
The soft SUSY parameters for NMSSM are also given.  The relevant numerical results are discussed in Sec \ref{secIII}. Sec \ref{secIV} contains our conclusions.

\section{ Messenger-Matter Interactions From Kahler Potential\label{secII} }
  There are two possible ways to deflect the AMSB trajectory with the presence of messengers, by pseudo-moduli field or
holomorphic terms (for messengers) in the Kahler potential. In our previous work\cite{Fei:1508.01299}, messenger-matter interactions are introduced
in the superpotential involving the pseudo-moduli field. We find that messenger-matter interactions can also be consistently generated by holomorphic messenger-matter mixing terms in the Kahler potential. In order to show the most general features of this approach, we discuss the relevant soft parameters in the framework of NMSSM.

As mentioned previously, the low energy soft SUSY breaking spectrum of NMSSM obtained from typical SUSY breaking mechanism, such as GMSB and ordinary extended (d)AMSB,
is always bothered by the requirement to achieve successful electroweak symmetry breaking(EWSB), rendering the model building non-trivial\cite{NMSSM:GMSB}.
In fact, EWSB within NMSSM in general requires a large VEV for the singlet. This prefers a negative $m^2_S$ and/or large $A$-terms for the singlet superpotential
interactions, $A_\la$ and $A_\ka$\cite{NMSSM:AMSB}. AMSB always predicts a positive value for $m^2_S$ and $A_\la,A_\ka$ can be sizable only if
$\la,\ka$ are also large which would induce a larger $m^2_S$, suppressing the singlet VEV.
Such difficulty can possibly be ameliorated in AMSB-type scenarios with enhanced trilinear couplings.
We find that phenomenological interesting NMSSM spectrum can be successfully generated by introducing holomorphic messenger-matter mixing terms in the Kahler potential.

We propose to introduce new holomorphic term in the Kahler potential in addition to ordinary canonical Kahler kinetic terms  and $Z_3$ invariant NMSSM superpotential
\beqa
{K_{h}}&\supseteq& {\phi^\da\phi}\left[\sum\limits_{i=1}^2c_{S,i} T S_i +c_P \tl{P}P+c_Q \tl{Q}Q+\sum\limits_{m=1}^3 \( c_{m,a}^P \tl{P}_{m,a} P+c_{m,a}^Q \tl{Q}Q_{m,a}\)\right]+h.c. ,\nn\\
 {W}&=&{\phi^3}\[W_M+\tl{\la} S_1 H_u H_d+\f{1}{3}\tl{\ka} S_1^3+W_{\overline{MSSM}}\]~,\nn\\
W_M&=&\sum\limits_{a=1,2,3}\la_P S_1 \tl{P}_{m,a} P+\la_Q S_1 \tl{Q} Q_{m,a} ~.
\eeqa
Here $\phi$ is the conformal compensator field and $S_1,S_2,T$ are gauge singlet superfields; $\tl{P}_{m,a}$ and $Q_{m,a}$ are the standard model matter superfields in the $\bar{\bf 5}$ and ${\bf 10}$ representations of SU(5) with $'a=1,2,3'$ the family index.
Additional vector-like messengers in the $\tl{P},P$ (${\bf \bar{5}} \oplus {\bf {5}}$) and $\tl{Q},Q$ (${\bf \overline{10}\oplus {10}}$) representations of SU(5) are introduced to solve the tachyonic slepton problems  and at the same time deflect the ordinary AMSB trajectory.
In this paper, we assume $c_P,c_Q,c_{m,a}^{P,Q}$ are real and $c_{m,a}^{P,Q}\neq0$ only for $a=3$ for simply. Such mixing between the third generation fermions and additional
vector-like fermions always appear in new physics models, such as top (top-bottom ) seesaw model or extra dimension models. We should note that $c_{m,a}^{P,Q}\neq 0$ for the first two generations $a=1,2$ are also possible and such possibility will be discussed subsequently.
 Besides, the choice of superpotential $W_M$ is not unique. For example, we can adopt couplings between $S$ and messengers used in \cite{ph:0604256} or \cite{giudice:0706.3873} (with double messenger species).

With only $c_{m,3}^{P,Q}\neq0$, the holomorphic terms in the Kahler potential reduce to
\beqa\label{combine}
K_h \supseteq  \f{\phi^\da}{\phi} \[T \(\sum\limits_{i=1,2} c_{S,i} S_i\)+P \(c_P \tl{P}+ c_{m,3}^P \tl{P}_{m,3}\)
+\tl{Q}\(c_Q Q+ c_{m,3}^Q Q_{m,3}\)\]+h.c.,
\eeqa
after rescaling each superfield with the compensator field $\phi$, namely  $\Phi \ra \phi \Phi$.
 With the F-term VEVs of compensator $\phi=1+F_\phi\theta^2$, we have the potential for the singlets
\beqa
\label{kahlersoft}
V \supseteq c_{S,i}|F_\phi|^2 T S_i+|F_\phi|^2(\sum\limits_{i} c_{S,i}^2) |T|^2 +|F_\phi|^2 |(\sum\limits_{i}c_{S_i} S_i )|^2+\cdots,
\eeqa
where the first term are obtained by picking out the $F_\phi$ terms.
  We thus arrive at the mass matrix for the scalar components $T,S_1,S_2$
 \beqa
 \label{tmatrix}
(~T,~S_1^*,~S_2^*) \left(\bea{ccc}
 (c_{S,1}^2+c_{S,2}^2)|F_\phi|^2&c_{S,1}|F_\phi|^2&c_{S,2}|F_\phi|^2\\ c_{S,1}|F_\phi|^2&c_{S,1}^2|F_\phi|^2&c_{S,1}c_{S,2}|F_\phi|^2\\c_{S,2}|F_\phi|^2&c_{S,1}c_{S,2}|F_\phi|^2&c_{S,2}^2|F_\phi|^2
 \eea \right)\left(\bea{c}T^*\\S_1\\S_2\eea\right)
\eeqa
It can be seen that the  mass matrix has vanishing determinant.

We can define $c_{S}\equiv \sqrt{c_{S,1}^2+c_{S,2}^2}$
and redefine the fields
\beqa
\tl{S}_0 &\equiv& \f{1}{c_S}\(c_{S,1} S_1+c_{S,2} S_2\)~,~~~~
\tl{S}_1 \equiv \f{1}{c_S}\(-c_{S,2} S_1+c_{S,1} S_2\)~,
\eeqa
The mixing angle can be given as
\beqa
\cos(-\theta)=\f{c_{S,1}}{\sqrt{c_{S,1}^2+c_{S,2}^2}}~, ~~~~\sin(-\theta)=-\f{c_{S,2}}{\sqrt{c_{S,1}^2+c_{S,2}^2}}~.
\eeqa
A minus sign for the angle is kept for future convenience.
The zero eigenvalue of the scalar mass matrix corresponds to the combination $\tl{S}_1$. The fermionic component of $\tl{S}_1$, which is orthogonal to $\tl{S}_0$, can also be seen to be massless from the Kahler potential. The non-vanishing mass eigenstates for scalar matrix are given by
\beqa
{\cal L}&\supseteq& -\f{(c_S^2-c_S)}{2c_S^2}F^2_\phi \left|-c_S T+c_{S,1}S_1^*+c_{S,2}S_2^*\right|^2
                  -\f{(c_S^2+c_S)}{2 c_S^2}F^2_\phi  \left|c_S T+c_{S,1}S_1^*+c_{S,2}S_2^*\right|^2~,\nn\\
       &=&   -{(c_S^2-c_S)} F^2_\phi \left|\f{-T+\tl{S}_0^*}{\sqrt{2}}\right|^2 -{(c_S^2+c_S)} F^2_\phi \left|\f{T+\tl{S}_0^*}{\sqrt{2}}\right|^2~.
\eeqa

Such expressions are analog to that of the GMSB with $T,\tl{S}_0$ the messenger-like fields. After integrated out the messengers, we can obtain their contributions to the low energy soft SUSY breaking spectrum. $S_1$ can be written as the combination
\beqa
S_1=\cos\theta \tl{S}_0+\sin\theta \tl{S}_1~,
\eeqa
which, after substituting into the superpotential, can lead to couplings between the massless fields $\tl{S}_1$ and heavy massive messenger-type fields $\tl{S}_0$.

Similar to the gauge singlet case, we can define the massive combinations $\bar{K}_1,K_1$ for $P,Q$ within eqn(\ref{combine}) and their massless orthogonal combinations $\bar{K}_2,K_2$ as
 \beqa
 \label{barK}
 \bar{K}_1 &\equiv& \f{1}{\sqrt{c_P^2+(c_{m,3}^P)^2 }} \[c_P \tl{P}+  c_{m,3}^P \tl{P}_{m,3}\],
  \bar{K}_2 \equiv \f{1}{\sqrt{c_P^2+(c_{m,3}^P)^2 }} \[-c_{m,3}^P \tl{P}+  c_P \tl{P}_{m,3}\],\nn\\
 K_1 &\equiv& \f{1}{\sqrt{c_Q^2+(c_{m,3}^Q)^2 }}\[c_Q Q+  c_{m,3}^Q Q_{m,3}\],
 K_2 \equiv \f{1}{\sqrt{c_Q^2+(c_{m,3}^Q)^2 }}\[-c_{m,3}^Q Q+  c_Q  Q_{m,3}\].
 \eeqa
So $\tl{P}_{m,3}(Q_{m,3})$ can be written as the combination of massive state $\bar{K}_1(K_1)$ and massless state $\bar{K}_2(K_2)$
\beqa
\tl{P}_{m,3}=\sin\psi_1\bar{K_1}+\cos\psi_1\bar{K}_2,~Q_{m,3}=\sin\psi_2{K_1}+\cos\psi_2{K}_2,
\eeqa
with
\beqa
\tan\psi_1=c_{m,3}^P/c_P,~~~~ \tan\psi_2=c_{m,3}^Q/c_Q~.
\eeqa

 The SUSY breaking effects from compensator F-term VEVs can be taken into account by introducing a spurion superfields $R$
\beqa
W=\int d^2 \theta  \left( c_P  R \tl{P}P+ c_Q R \tl{Q} Q +\cdots \right),
\eeqa
with the spurion VEV as
\beqa
R\equiv M_R+\theta^2 F_R= F_\phi(1-\theta^2 F_\phi)~,
\eeqa
which gives the deflection parameter
\beqa
d\equiv\f{F_R}{M_R F_\phi}-1=-2.
\eeqa
The spurion messenger-matter interactions will affect the AMSB RGE trajectory after integrating out the heavy modes.

  The superpotential involving the matter and singlet superfields at the $F_\phi$ scale are given as
 \beqa
W &\supseteq& \tl{\la} S_1 H_u H_d+\f{1}{3}\tl{\ka} S_1^3+\sum\limits_{a=1,2,3}\[\la_P S_1 \tl{P}_{m,a} P+\la_Q S_1 \tl{Q} Q_{m,a}\]+y_{ij}^{\bf \bar{5}}\tl{P}_{m,i}Q_{m,j}\bar{H}_{\bf \bar{5}}~\nn\\
&+& {y}^{\bf 5}_{ij}Q_{m,i}Q_{m,j}H_{\bf 5}+\sqrt{c_P^2+(c_{m,3}^P)^2}\bar{K}_1 P R+\sqrt{c_Q^2+(c_{m,3}^Q)^2 }\tl{Q} K_1 R ~,\nn\\
&\supseteq& (\cos\theta \tl{S}_0+\sin\theta \tl{S}_1){\tl{\lambda}} H_u H_d+ \f{\tl{\kappa}}{3}(\cos\theta \tl{S}_0+\sin\theta \tl{S}_1)^3\nn\\
&+& \[{\la_P}\( \sin\psi_1\bar{K}_1+\cos\psi_1 \bar{K}_2\) P+{\la_Q}\tl{Q}\(\sin\psi_2 {K}_1+\cos\psi_2 K_2\)\](\cos\theta \tl{S}_0+\sin\theta \tl{S}_1),\nn\\
&+&\sum\limits_{a=1,2}\({\la_P} \tl{P}_{m,a} P+{\la_Q}\tl{Q} Q_{m,a}\)(\cos\theta \tl{S}_0+\sin\theta \tl{S}_1)+\cdots~.
\eeqa
The terms containing the ${\bf 10,\bar{5}}$ representations of SU(5) reduce to the sum of their components below the GUT scale and should be understood as the abbreviation of this sum at low energy
\beqa
&&\sum\limits_{a=1,2,3}\[\la_P S_1 \tl{P}_{m,a} P+\la_Q S_1 \tl{Q}Q_{m,a} \]+y_{ij}^{\bf \bar{5}}\tl{P}_{m,i}Q_{m,j}\bar{H}_{\bf \bar{5}}+ {y}^{\bf 5}_{ij}Q_{m,i}Q_{m,j}H_{\bf 5}\\
\Longrightarrow && \sum\limits_{a=1,2,3}S_1\[ \la_{D,a} (D_{L,a}^c) D+\la_{L,a} (L_{L,a}) L+\la_{Q,a}(Q_{L,a}) Q+\la_{U,a}(U_{L,a}^c) U+\la_{E,a}(E_{L,a}^c) E\]\nn\\
&&~~~+y_{ab}^U (Q_{L,a})(U_{L,b}^c) H_u+y_{ab}^D (Q_{L,a})(D_{L,b}^c) H_d
+ y_{ab}^E (L_{L,a}) (E_{L,b}^c) H_d~.\nn
\eeqa
The same holds for terms containing $K_a,\bar{K}_a$. Besides, terms involving the triplet components of $H,\bar{H}$ are integrated out by assuming proper doublet-triplet
splitting mechanism to generate heavy triplet Higgs masses.

After integrated out all the heavy modes including $T,\tl{S}_0;\bar{K}_1,K_1;P,\tl{Q}$ etc, the low energy theory will reduce to NMSSM.
The effects of integrating out the messengers can be taken into account
by using Giudice-Rattazi's wavefunction renormalization approach\cite{GMSB:wavefunction}.
The messenger threshold $M_R$ can be further promoted to the other chiral spurion field $X$ with $M_R=\sqrt{X^\da X}$.
The superfield $\tl{S}_1$ will act as the singlet $S$ appearing in ordinary NMSSM superpotential and $\bar{K}_2,K_2$ as the third generation superfields.
Note that there is a scaling of various couplings appeared in NMSSM
  \beqa
  \label{ytyb}
  \la=\tl{\la}\sin\theta,~\ka= \tl{\ka}\sin^3\theta,~y_{b,\tau}=y_{33}^{D,L}\cos\psi_1\cos\psi_2,~y_t=y_{33}^U\cos^2\psi_2~.
  \eeqa
   In the subsequent studies, we must ensure that correct scaled (or unscaled) couplings are used in the expressions.  New flavor dependent interactions involving the messengers from $y_{ij}^{D,L}\tl{P}_{m,i}Q_{m,j}\bar{H}_{\bf \bar{5}}$ and ${y}^U_{ij}Q_{m,i}Q_{m,j}H_{\bf 5}$ are not dangerous because
   these new flavor dependent interactions are aligned with the MSSM Yukawa coupling. Diagonalizing the MSSM Yukawa couplings will simultaneously diagonalize these additional Yukawa couplings.

We should briefly discuss the most general case with $c^P_{m;1,2}\neq 0$. Similar to eqn.(\ref{barK}), we can define the mixing matrix $U^P_{AB}$ with the indices $A,B=(0,a)~~[a=1,2,3]$
\beqa
\(\bea{c}\bar{K}_0\\ \bar{K}_1\\ \bar{K}_2\\\bar{K}_3 \eea\) &\equiv& \f{1}{\sqrt{c_P^2+\sum\limits_{a}(c_{m,a}^P)^2}} \(\bea{cccc}c_P&c^P_{m,1}&c^P_{m,2}&c^P_{m,3}\\-c^P_{m,1}&c_P&c^P_{m,3}&-c_{m,2}^P\\ -c^P_{m,2}&-c^P_{m,3} &c_P&c_{m,1}^P\\-c^P_{m,3}&c_{m,2}^P&-c_{m,1}^P&c_P\eea\)
\(\bea{c}\tl{P}\\ \tl{P}_{m,1}\\\tl{P}_{m,2}\\\tl{P}_{m,3} \eea\)~.
\eeqa
within which the three orthogonal combinations $\bar{K}_a (a=1,2,3)$ are determined up to an arbitrary rotation under $SO(3)$ transformation ${\cal O}^P$.
The scalar components of the three combinations $\bar{K}_a (a=1,2,3)$ correspond to the massless eigenvalues of the $5\times 5$ sfermion mass matrix for $(P^*,\tl{P},\tl{P}_1^m,\tl{P}_2^m,\tl{P}_3^m)$ that can be identified to be the three generation squark/sleptons in MSSM. Similar conclusions hold for $Q^m_a$. 
Besides, the relations between $y_t,y_b,y_\tau$ and $y^U_{33},y^D_{33},y^L_{33}$ will be non-trivial and depends on various parameters
in the mixing matrix
   \beqa
   \label{rotate}
  \({\cal O^Q} y_{SM}^U{\cal O^Q}\)_{ab}= (U^Q)^{-1}_{a c} y^U_{cd} (U^Q)^{-1}_{db}~,\nn\\
   \({\cal O^P} y_{SM}^D{\cal O^Q}\)_{ab}= (U^P)^{-1}_{a c} y^D_{cd} (U^Q)^{-1}_{db}~,\nn\\
   \({\cal O^P} y_{SM}^L{\cal O^Q}\)_{ab}= (U^P)^{-1}_{a c} y^L_{cd} (U^Q)^{-1}_{db}~.
   \eeqa
   with
   \beqa
   (U^P)^{-1}=\f{(U^P)^T}{c_P^2+\sum\limits_{a}(c_{m,a}^P)^2}~\nn.
   \eeqa
 After fixing the rotation matrix ${\cal O}^P$ and ${\cal O}^Q$, we can obtain precisely the relations between SM Yukawa couplings and the couplings in the superpotential. For new flavor dependent interactions involving one messengers, such as $(y^U)^\pr_{0 b}\bar{K}_{Q;0}\bar{K}_{U_L^c;b}H_u$, the coupling can be seen to satisfy
 \beqa
 (y^U)^\pr_{0 c}{\cal O^Q}_{c b}&=& (U^Q)^{-1}_{0 c} y^U_{cd} (U^Q)^{-1}_{db}~,\nn\\
  (y^U)^\pr_{0 c}&=&(U^Q)^{-1}_{0 a} U^Q_{a e}{\cal O}^Q_{e d}(y^U_{\bf SM})_{d c}~.
 \eeqa
 by combining with eqn(\ref{rotate}).
 So the flavor constraints are no-dangerous if
 $U^Q_{ae} {\cal O}_{ed}^Q$ is diagonal so that $(y^U)^\pr$ can align to $y^U_{\bf SM}$ or the coefficients of the new Yukawa couplings
 \beqa
  (U^Q)^{-1}_{0 a}= -\f{c^Q_{m,a}}{c_Q^2+\sum\limits_{a}(c_{m,a}^Q)^2}~,
  \eeqa
   are small. The SO(3) rotation matrix can be parameterized by three Euler angles, so the three diagonal elements of the diagonalized $U^Q_{ab} {\cal O}_{bc}^Q$ matrix can match the three rotation freedom. Therefore, it is possible to align $(y^U)^\pr$ and $y^U_{SM}$ by proper chosen ${\cal O}^Q$ so that we need not worry too much about the flavor constraints. On the other hand, if we insist on small $c_{1,2}^{P,Q}$, such small numbers can be the consequence of suppressions from additional Froggatt-Nielsen\cite{FN-1,FN-2} type mechanism with additional horizontal flavor symmetry.


After integrating out the messenger fields, the wavefunction will depend on the messenger threshold set by the spurion superfield $R$.
 The soft gaugino masses are given at the messenger scale by
\beqa
M_{i}(M_{mess})&=& g_i^2\(\f{F_\phi}{2}\f{\pa}{\pa \ln\mu}-\f{d F_\phi}{2}\f{\pa}{\pa \ln |X|}\)\f{1}{g_i^2}(\mu,|X|,T)~,
\label{sgaugino}
\eeqa
with
\beqa
\f{\pa}{\pa \ln |X|} g_i(\al; |X|)=\f{\Delta b_i}{16\pi^2} g_i^3~,
\eeqa

Because of the non-renormalization of the superpotential, the trilinear soft terms will be determined by the wavefunction normalization as
\beqa
A_0^{ijk}\equiv \f{A_{ijk}}{y_{ijk}}&=&\sum\limits_{i}\(-\f{F_\phi}{2}\f{\pa}{\pa\ln\mu}+{d F_\phi}\f{\pa}{\pa\ln X}\) \ln \[Z_i(\mu,X,T)\]~,\nn\\
&=&\sum\limits_{i} \(-\f{F_\phi}{2} G_i^- +d F_\phi\f{\Delta G_i}{2}\)~.
\eeqa
The anomalous dimension are expressed in the holomorphic basis\cite{shih} as
\beqa
G^i\equiv \f{d Z_{ij}}{d\ln\mu}\equiv-\f{1}{8\pi^2}\(\f{1}{2}d_{kl}^i\la^*_{ikl}\la_{jmn}Z_{km}^{-1*}Z_{ln}^{-1*}-2c_r^iZ_{ij}g_r^2\).
\eeqa
with $\Delta G\equiv G^+-G^-$ the discontinuity across the messenger threshold.
Here $G^+(G^-)$ denotes the value above (below) the messenger threshold, respectively.

The soft SUSY breaking scalar masses are given by
\beqa
m^2_{soft}&=&-\left|-\f{F_\phi}{2}\f{\pa}{\pa\ln\mu}+d F_\phi\f{\pa}{\pa\ln X}\right|^2 \ln \[Z_i(\mu,X,T)\]~,\\
&=&-\(\f{F_\phi^2}{4}\f{\pa^2}{\pa (\ln\mu)^2}+\f{d^2F^2_\phi}{4}\f{\pa}{\pa(\ln |X|)^2}
-\f{d F^2_\phi}{2}\f{\pa^2}{\pa\ln|X|\pa\ln\mu}\) \ln \[Z_i(\mu,X,T)\],\nn
\label{sscalar}
\eeqa
 The dAMSB soft scalar masses can be divided into several parts,
 namely the gauge-anomaly interference part, the pure gauge mediation part as well as the ordinary anomaly mediation part.

\section{Numerical Results\label{secIII}}
From the previous general formulas for soft SUSY breaking parameters, we can obtain the analytical expressions in our scenario at the scale $F_\phi$
after integrating out the messenger fields. Some of the lengthy expressions are given explicitly in the appendix.
We can see that new contributions will be given to the trilinear couplings $A_\ka.A_\la,m_S^2$ which could be helpful to trigger EWSB in NMSSM.

   We use NMSSMTools5.1.2\cite{NMSSMTOOLS} to scan the whole parameter space. The free parameters of our inputs are chosen to satisfy
 \beqa
&&~~~~10 {\rm TeV} < F_\phi< 1000{\rm TeV}~,~~~~~~0.1<\cot\theta,\tan\psi_1,\tan\psi_2< 10~,~\nn\\
&&~~~~~~~~~~0< \la_{D,a},\la_{L,a},\la_{Q,a},\la_{U,a},\la_{E,a}<\sqrt{4\pi}~,~~~~~~~0<\la,\ka<0.7~,~~
 \eeqa
 with the range of the mixing angle $-\pi/2<\theta,\psi_1,\psi_2<\pi/2$ and we require that $\la^2+\ka^2\lesssim 0.7$ to satisfy the perturbative bounds..
The soft SUSY mass $m_{H_u}^2,m_{H_d}^2,m_S^2$ can be recast into $\mu,\tan\beta,M_Z^2$ by the minimization conditions of the scalar potential.
Usually, the parameter $M_A$ can be used to replace $A_\ka$ by
  \beqa
  M_A^2=\f{2\mu_{eff}}{\sin2\beta}B_{eff}~, ~~\mu_{eff}\equiv\la\langle s\rangle~,~~~B_{eff}=(A_\la+\ka \langle s\rangle).
  \eeqa
  We note that $\kappa$ is a free parameter while the $\tan\beta\equiv v_u/v_d$ is not. This choice is different to ordinary numerical setting from top-down
in which $\tan\beta$ is free while $\kappa$ is a derived quantity\cite{ellwanger:0612134}.
Such choice is convenient for those predictable NMSSM models with a UV completion.
Firstly, we need to guess a value for $\tan\beta$ to obtain the relevant Yukawa $y_t,y_b$ couplings at the electroweak(EW) scale.
After renormalization group equation(RGE) evolution of $g_i,y_t,y_b$ couplings from EW scale to the messenger scale as the theory inputs,
the whole soft SUSY breaking parameters at the messenger scale can be obtained.
Low energy $\tan\beta$ can be obtained iteratively from the minimization condition of the Higgs potential\cite{Fei:1704.05079}.
 The purpose of deflection in AMSB is to solve the notorious tachyonic slepton problem. So non-tachyonic slepton should be obtained at the EW scale.

In our scan, we also impose the following collider constraints
\bit
\item (i)  The lower bounds from current LHC constraints on SUSY particles\cite{atlas:gluino,atlas:stop}. In particular,
        the gluino mass is bounded $m_{\tl{g}} \gtrsim 1.85\sim 2.0 $ TeV from a search for gluino pair production,
        assuming $\tl{g}\ra q\bar{q}\chi_1^0$ with a massless LSP and decoupled squarks; or $m_{\tl{g}} \gtrsim 1.96\sim 2.05 $ TeV,
        assuming $\tl{g}\ra t\bar{t}\chi_1^0$. The squark $m_{\tl{t}_1}\gtrsim 0.95\sim 1.05$ TeV for light third generation sfermions, and
        $m_{\tl{q}}\gtrsim 1.6$ TeV for the first two generations.

\item  (ii) Flavor constraints from the rare decays of B meson. We adopt the recent experimental results\cite{B-physics}:
  \beqa
  && 0.85\tm 10^{-4} < Br(B^+\ra \tau^+\nu) < 2.89\tm 10^{-4}~,\nn\\
  && 2.99\tm 10^{-4} < Br(B_S\ra X_s \gamma) < 3.87\tm 10^{-4}~,\nn\\
  && 1.7\tm 10^{-9} < Br(B_s\ra \mu^+ \mu^-) < 4.5\tm 10^{-9}~,
   \eeqa

\item  (iii) The CP-even component $S_2$ in the Goldstone-$'eaten'$ combination of $H_u$ and $H_d$ doublets corresponds to the SM-like Higgs.
 The $S_2$ dominated CP-even scalar should lie in the combined mass range: $122 {\rm GeV}<M_h <128 {\rm GeV}$ from ATLAS and CMS data.
 Note that the uncertainty is 3 GeV instead of default 2 GeV in NMSSMTools because large $\lambda$ may induce additional
 ${\cal O}(1)$ GeV correction to $m_h$ at two-loop level\cite{NMSSM:higgs2loop}.

\item (iv) The EW precision observables\cite{precision} and the lower bounds for the neutralino and chargino masses, including the invisible decay bounds for $Z$ boson. The most stringent LEP bounds require $m_{\tl{\chi}^\pm}> 103.5 {\rm GeV}$ and the invisible decay width $\Gamma(Z\ra \tl{\chi}_0\tl{\chi}_0)<1.71~{\rm MeV}$, which is consistent with the $2\sigma$ precision EW measurement $\Gamma^{non-SM}_{inv}< 2.0~{\rm MeV}$\cite{LEP}.

\eit

  A possible hint of new physics beyond the SM is the muon $g-2$ anomaly.
 The E821 experimental result of the muon anomalous magnetic moment at the Brookhaven AGS \cite{Mg-2:collaboration} is given by
 \beqa
a_\mu^{\rm expt} =116592089(63)\times 10^{-11}~,
 \eeqa
which is larger than the SM prediction\cite{Mg-2:SM}
\beqa
a^{\rm SM}_\mu =116591834(49)\times 10^{-11}~.
\eeqa
We adopt the conservative estimation $4.7\tm 10^{-10}\lesssim\Delta a_\mu\lesssim52.7\tm 10^{-10}$ in our scenario.
 If possible, we would like our theory to explain such muon g-2 anomaly.

To address the strong CP problem in SUSY, the axino, which is the fermionic superpartner
of axion, can be predicted. If the existence of axino could be confirmed by experiments, the two theoretical hypotheses, which are designed
to solve two respective hierarchy problems: the strong CP problem by a very light axion and the gauge hierarchy problem by SUSY, could be validated.

 The SUSY version of the Kim-Shifman-Vainshtein-Zakharov (KSVZ)\cite{KSVZ} axion model,
 which introduces a SM singlet and a pair of extra vector-like quarks that carry $U(1)_{PQ}$ charges
 while keeps the SM fermions and Higgs fields neutral under $U(1)_{PQ}$ symmetry,
 could be introduced in NMSSM with additional singlets and messengers.
The decoupling theorem\cite{AMSB:decouple} of anomaly mediation, which states that pure mass threshold will not deflect the AMSB trajectory,
guarantee that such new messengers, with their masses determined by the singlet VEVs between $10^{10}{\rm GeV}\lesssim \langle S\rangle \lesssim 10^{12}$ GeV
(to be compatible with the experimental bound on $f_a$\cite{axion:PR}) will not change the original AMSB predictions upon $F_\phi$.

The axino mass, which is strongly model dependent, can be much smaller or much larger than the SUSY scale.
Some models predict $m_{\tl{a}}$ of the order of gravitino mass $m_{3/2}$\cite{axion:PR}
while the spontaneously broken global SUSY predict it to lie of order $m_{\tl{a}} \sim {\cal O}(M^2_{SUSY}/f_a)$\cite{light:axino}.
So one usually treats axino mass as a free parameter and assumes the axino interactions to be given by the $U(1)_{PQ}$ symmetry.
If axino is heavy,  the lightest neutralino will be DM candidate; if the axino is the LSP and acts as DM particle,
colored or charged NLSP can decay into axino and non-thermally generate the observed axino DM relic density.

Depending on the nature of the LSP, we have the following discussions from our numerical results
\begin{itemize}
\item  Scenario I:  The lightest neutralino $\chi_1^0$ as the LSP
\begin{figure}[htb]
\label{scI}
\begin{center}
\includegraphics[width=2.9in]{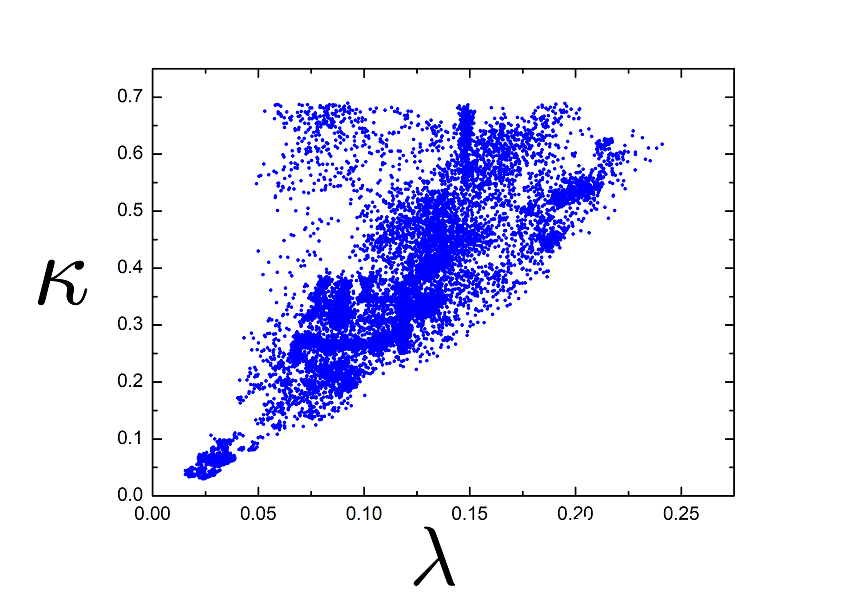}
\includegraphics[width=2.9in]{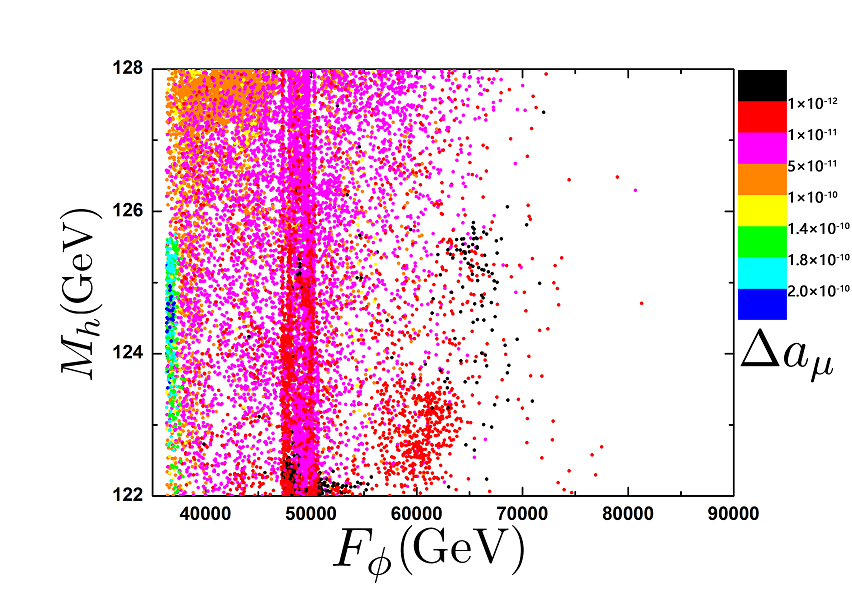}\\
\includegraphics[width=2.9in]{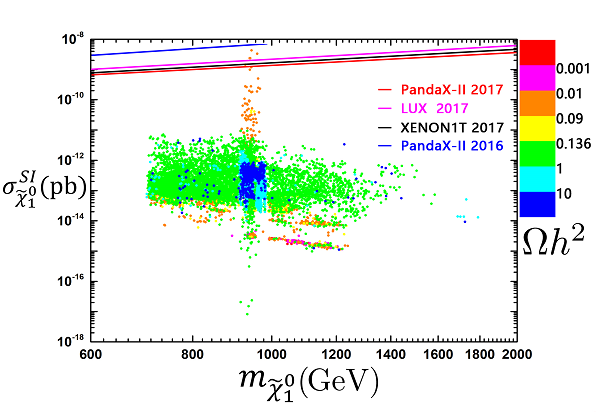}
\includegraphics[width=2.9in]{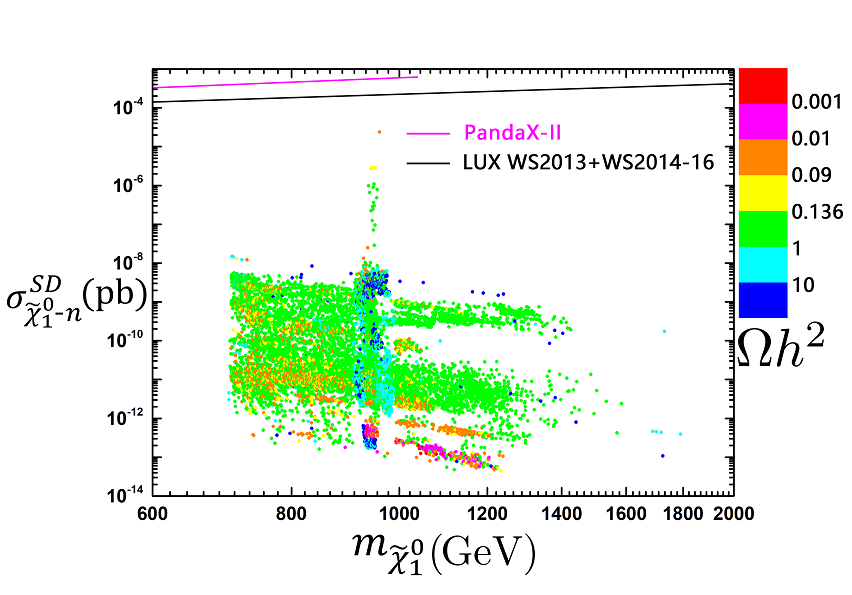}\\
\end{center}
\caption{Numerical scan for scenario I with neutralino LSP. All points can satisfy the collider constraints from (i) to (iv).
The allowed values for $\la$ vs $\ka$, which are derived from EWSB conditions of NMSSM, are given in the upper left panel;
in the upper right panel, the allowed ranges of $F_\phi$ vs $m_{Higgs}$ are shown explicitly. The corresponding values of SUSY contributions to
$\Delta a_\mu$ are also shown; in the lower left panel and lower right panel, the Spin-Independent and Spin-Dependent direct detection cross sections for neutralino dark matter are given, respectively. The exclusion lines from from LUX and PANDAX experiments are also shown.
}
\end{figure}

  Successful EWSB conditions impose stringent constrains on the input parameters of NMSSM,
especially when such low energy inputs are determined by a UV-completed theory. Random scan in this case
indicates that still some points can survive the EWSB conditions by leading to a iteratively stable value of $\tan\beta$.
The allowed range for characteristic NMSSM parameters $\ka$ and $\la$ can be seen in the upper left panel of Fig 1.

 Besides, the discovered 125 GeV Higgs can be successfully interpreted in our scenario.
 The mass scales of the soft spectrum are uniquely determined by the gravitino mass $F_\phi$ in AMSB-type scenarios.
 The plot of $F_\phi$ versus Higgs boson mass is shown in the upper right panel of Fig 1.
 In the NMSSM, the SM-like Higgs can be pushed to 125 GeV by additional tree-level contributions for relatively large $\la$ and small $\tan\beta$.
  If the lightest CP-even scalar is mostly singlet-like, mixing between it and the SM-like Higgs can lead to an increase of the SM-like Higgs mass.
 Although the possibility of SM-like Higgs being the second lightest CP-even scalar is attractive, we find that almost all the survived points predict the lightest CP-even scalar as the 125 GeV Higgs.

Numerical results indicate that the muon $g-2$ anomaly cannot be explained in this scenario.
The SUSY contributions to $\Delta a_\mu$ are shown in different color in the upper right panel of Fig 1.
This result can be understood because the required $\Delta a_\mu$ can be achieved only if
the relevant sparticles( $\tl{\mu},\tl{\nu}_\mu, \tl{B},\tl{W}, \tl{H}$) are lighter than $600\sim 700$ GeV for $\tan\beta\sim 10$ in MSSM\cite{Mg-2:SUSY}.
The inclusion of singlino in NMSSM will not give sizable contributions to $\Delta a_\mu$
because the coupling of singlino to MSSM sector is suppressed.
Although the two loop contributions involving the Higgs is negligible in SM, the new higgs bosons in NMSSM could have an important impact on $a_\mu$ if the lightest neutral CP-odd Higgs scalar is very light\cite{NMSSM:g-2}.
As noted there, positive two-loop contribution is numerically more important for a light CP-odd Higgs heavier than 3 GeV and
the sum of both one-loop and two-loop contributions is maximal around $m_{a_1}\sim 6$ GeV.
In our scenario, the lightest CP-odd Higgs $a_1$ is not light enough to give sizable contributions to $\Delta a_\mu$.
So the main contributions are similar to that in the MSSM and are not large enough to account for the discrepancy.

 In the case of $\chi_1^0$ as the LSP, the dark matter(DM) relic density is required to satisfy the bounds set by
the Planck data\cite{Planck} in combination with the 5$\sigma$ WMAP data\cite{WMAP}:
 $$0.0913\lesssim \Omega_{DM}\lesssim 0.1363.$$
Numerical scan indicates that the LSP is almost pure Bino-like.
This fact can be understood from the gaugino ratio at the $F_\phi$ scale $M_1:M_2:M_3=14.6:9:5$ which will lead to the ratio at the EW scale
$M_1:M_2:M_3\approx 14.6:18:30$. The Higgsino components within the LSP are tiny and the effective $\mu$ parameter is relatively high in our scenario.

 It is well known that the Bino LSP will lead to over-abundance of DM unless co-annihilation with sfermions or s-channel resonance increase the annihilation rates.
In our scenario, the Bino-like DM always co-annihilates with the lightest stau or sbottom and leads to the right DM relic density.
The mass differences between $\tl{\tau}_1(\tl{b}_1)$ NLSP and $\tl{\chi}_1^0$ LSP are typically of order ${\cal O}(GeV)$.


Interactions between Bino dark matter and the nucleons are primarily mediated by t-channel scalar Higgses or by s-channel squarks.
The Higgs exchange diagrams dominate the Spin-Independent (SI) cross section of $\chi_1^0$ on nucleon for sufficiently heavy squarks.
The Spin-Dependent(SD) cross section, which is dominated by the Z-boson exchange diagram as long as the squarks are sufficiently heavy,
can provide a good probe of the gaugino and Higgsino parameters. Large SD cross sections can appear only if $\chi_1^0$ DM contains a large Higgsino component.
Although the SD direct detection cross section is generally larger than the SI cross section,
it is much more difficult to probe in the experiment, as the SD cross section does not scale directly with the mass of the nuclei.
  In fact, current bounds on the neutron SD cross section are less stringent by a factor of $10^6$.

The SI and SD direct detection results are shown in the lower right panel of Fig.1.
 In our scenario, the neutralino DM is almost pure Bino with very tiny Higgsino and Wino contents.
  Such (almost pure) Bino-like DM can survive the direct detection constraints from  LUX\cite{LUX2016} and PANDAX\cite{PANDAX} because of the suppressed Higgsino component.

 \item  Scenario II: The axino as the LSP

\begin{figure}[htb]
\label{scII}
\begin{center}
\includegraphics[width=2.9in]{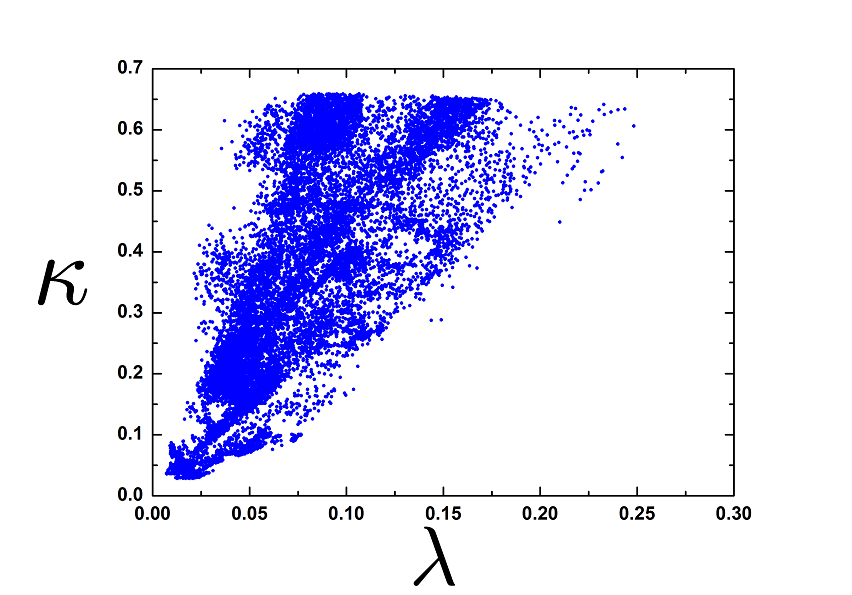}
\includegraphics[width=2.9in]{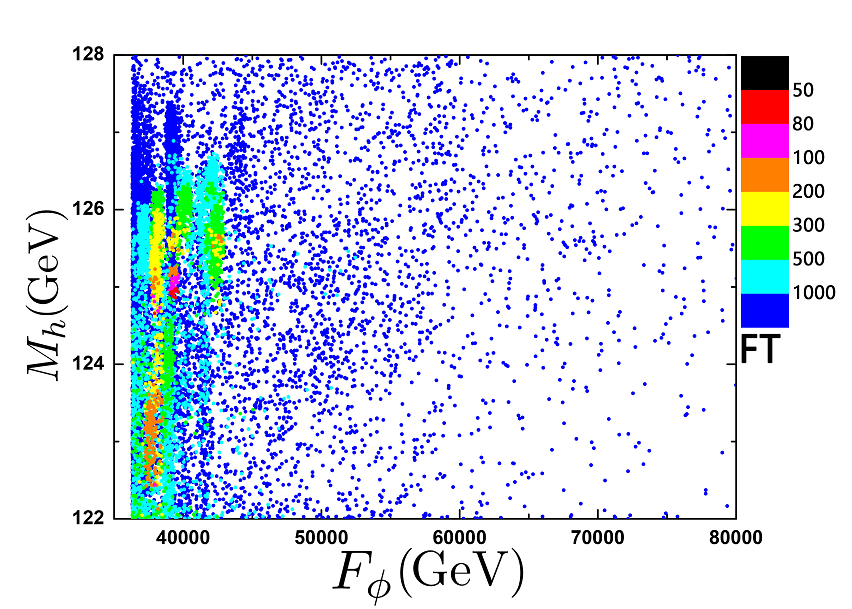}\\
\includegraphics[width=2.9in]{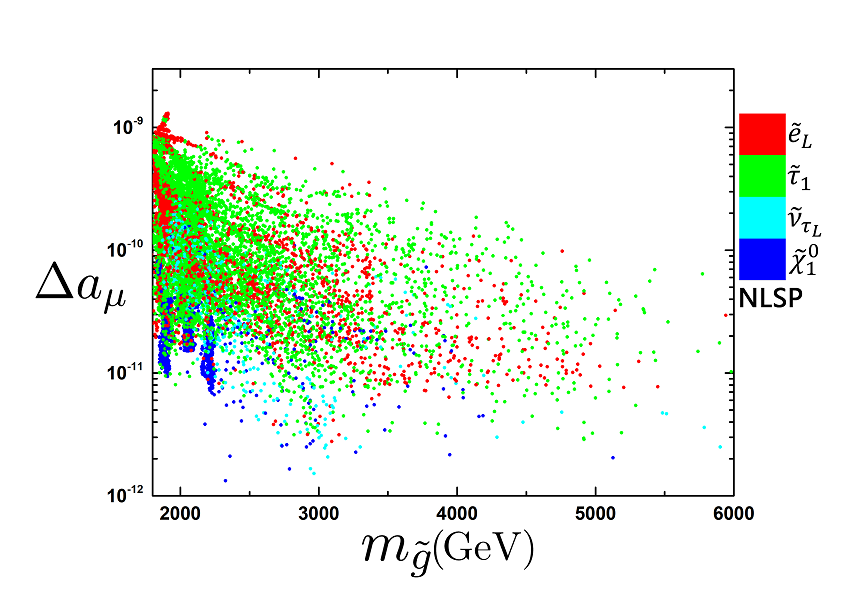}
\includegraphics[width=2.9in]{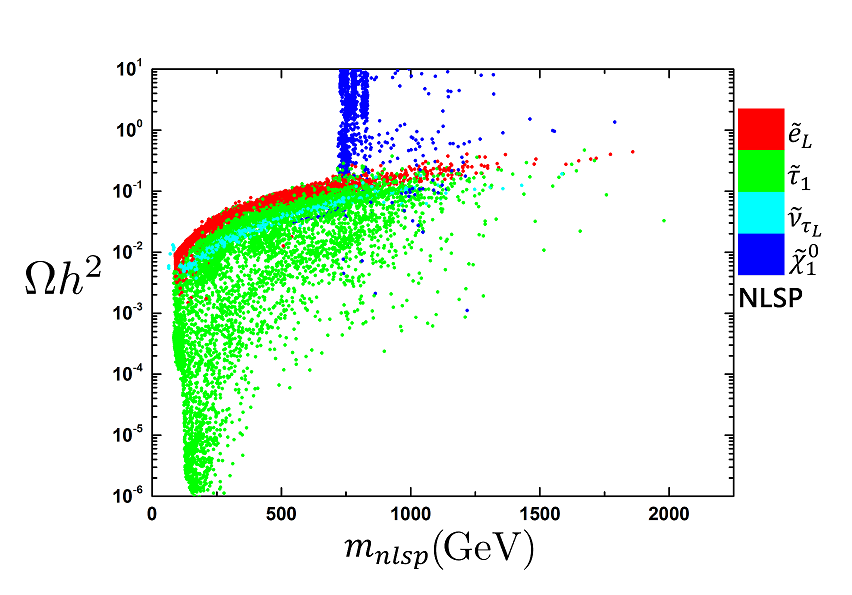}\\
\end{center}
\caption{ Numerical scan for scenario II with axino LSP. All points can satisfy the collider constraints (i) to (iv).
In the upper left panel, the allowed values for $\la$ vs $\ka$ are derived from EWSB condition of NMSSM;
in the upper right panel, the allowed ranges of $F_\phi$ vs $m_{Higgs}$ are shown explicitly. The involved Barbieri-Giudice fine tuning are also shown; in the lower left panel, the SUSY contributions to $\Delta a_\mu$ vs the gluino mass $m_{\tl{g}}$ are given for various NLSP($\tl{\tau}_1,\tl{e}_R,\tl{b}_1$ etc);
 The freeze-out relic density for NLSP particles (before later decaying into axino DM) are shown in the lower right panel.}
\end{figure}

 As mentioned before, the axino can be the LSP and act as the DM candidate.
 So the previously forbidden charged or colored LSP region can be revived.
 In some of the allowed parameter space,  numerical scan indicate that
 lightest ordinary supersymmetric particle(LOSP) will be the lightest sbottom, stau, tau-sneutrino etc.

 Even in this scenario, successful EWSB conditions can impose stringent constrains on the input parameters of NMSSM,
especially for characteristic NMSSM parameters $\ka$ and $\la$.
Their allowed range can be seen in the upper left panel of Fig 2.
The plot of $F_\phi$ versus Higgs boson mass is shown in the upper right panel of Fig 2.
We can see that the 125 GeV Higgs mass can also be easily accommodated in this scenario.

The muon $g-2$ discrepancy, however, prefer this scenario. The most stringent constraint, as noted in our previous papers\cite{Fei:1703.10894}, is the gluino mass bound.
 We can see that the SUSY contributions to $\Delta a_\mu$ can reach $14\tm 10^{-10}$
in the case of slepton($\tl{\tau}_1,\tl{e}_R$) LOSP with $m_{\tl{g}}\approx 1.8$ TeV. The mass bound of gluino can be understood as follows:
 as the whole low energy spectrum is determined by the value of $F_\phi$, the mass scales of $\tl{\mu}$ etc
determine the upper bound of $F_\phi$, which, on the other hand, sets a bound on gluino mass.
We find that the gluino mass is bounded to less than 2.2 TeV if we adopt the lower limit of the required $\Delta a_\mu$.
Such light gluino will soon be tested by the future LHC searching results.

The explanation of the muon $g-2$ discrepancy prefers the effective $\mu$ parameter to be light, which could reduce the fine tuning (FT) involved in our theory.
Besides, the stop mass can also be predicted to lie just upon the experimental bounds from recent LHC and be fairly natural\cite{rnaturalsusy}.
We show the Barberi-Giudice(BG) FTs\cite{BGFT} for the survived points in the figures.  We can see that the FT involved in this scenario can be as low as 50.

DM particle in this scenario is the axino from LOSP decay (neglecting the effects of saxion condensate\cite{saxion}).
As the non-thermally produced relic density of axino can be related to that of NLSP via
\beqa
\Omega_{\tl{a}}=\f{m_{\tl{a}}}{m_{NLSP}}\Omega_{NLSP}~,
\eeqa
the relic density of DM will not impose stringent constraints on our parameter space. We show the relic density of NLSP in Fig 2.
The thermal production of axino, which may give important contributions (or even dominate over the non-thermally production) to the observed relic density and be model dependent, will impose strong bound on the reheating temperature after inflation by requiring such production should not overclose the universe\footnote{It is noted in \cite{axino:squark} that, even though squarks are normally not the NLSP and remain in thermal equilibrium, axino yield from squark decay can dominate the abundance for $T_R\lesssim m_{\tl{q}}$ and large gluino mass. We assume the reheating temperature is higher than the squark masses.}. Relevant discussions on the reheating temperature and cosmological consequences of axino DM can be found in\cite{axino:reheating}.

The extremely weak interaction strength of axino makes its detection in DM direct detection experiments, as well as at collider experiments,
rather hopeless. The hint of the axino DM may show up from the properties of the LOSP.
In the case of axino LSP, the LOSP typically decays with a lifetime of less than one second and practically be stable inside the collider detector.
The (electrically) charged or colored  particle as a LOSP would appear as a stable particle inside the detector.
The injection of high-energetic hadronic and electromagnetic particles, produced from late decays of an NLSP to axino (with lifetime less than one second),
will not affect the abundance of light elements produced during Big Bang Nucleosynthesis(BBN).

\end{itemize}

\section{Conclusions\label{secIV}}
 We propose a new approach to generate messenger-matter interactions in deflected anomaly mediated SUSY breaking mechanism from typical holomorphic messenger-matter mixing terms in the Kahler potential. This approach is a unique feature of AMSB and has no analog in GMSB-type scenarios. New coupling strengths obtained from the scaling of the (already known) Yukawa couplings, such as $y_t\tan\psi_2,y_t\tan^2\psi_2,y_b\tan\psi_2\tan\psi_1,\ka\cot\theta,\ka\cot^2\theta$, always appear in this approach, which is a salient feature of this scenario.
 With messenger-matter interactions in deflected AMSB, we can generate a realistic soft SUSY breaking spectrum for next-to-minimal supersymmetric standard model(NMSSM). Successful electroweak symmetry breaking conditions, which is not easy to satisfy in NMSSM for ordinary AMSB-type scenario, can be satisfied in a large portion of parameter space in our scenarios. We study the relevant phenomenology for scenarios with (Bino-like) neutralino and axino LSP, respectively.
 In the case of Bino-like LSP, most of parameter space can survive the recent SI and SD DM direct detection experiment;
 in the case of axino LSP, the SUSY contributions to $\Delta a_\mu$ can possibly account for the muon g-2 discrepancy.
  The corresponding gluino masses (for those points that can explain the muon g-2 anomaly), which are found to below 2.2 TeV, could be tested soon at LHC.
\begin{acknowledgments}
 We are very grateful to the referee for efforts to improve our draft. This work was supported by the
Natural Science Foundation of China under grant numbers 11675147,11105124; by the Innovation Talent project of Henan Province under grant
number 15HASTIT017; by the Young Core instructor foundation of the Henan education department.
\end{acknowledgments}

\section*{Appendix: The Soft SUSY Breaking Parameters}
We list the relevant soft SUSY breaking parameters obtained from the analytical formulas (\ref{sgaugino}) to (\ref{sscalar}).

For gaugino masses, we have
\beqa
M_i=-F_\phi\f{\al_i(\mu)}{4\pi}\(b_i-(-2)\Delta b_i\)~,
\eeqa
with
\beqa
~(b_1~,b_2~,~b_3)&=&(\f{33}{5},~1,-3)~,~~\nn\\
\Delta(b_1~,b_2~,~b_3)&=&(~4,~4,~4).
\eeqa


The trilinear couplings are given as
\beqa
A_t&=&\f{F_\phi}{16\pi^2}\[\tl{G}_{y_t}-d \Delta \tl{G}_{y_t}  \]~,\\
A_b&=&\f{F_\phi}{16\pi^2}\[\tl{G}_{y_b}-d \Delta \tl{G}_{y_b} \]~,\nn\\
A_\tau&=&\f{F_\phi}{16\pi^2}\[\tl{G}_{y_\tau}-d \Delta\tl{G}_{y_\tau} \]~,\nn\\
A_\la &=&\f{F_\phi}{16\pi^2}\left\{\tl{G}_{\la}-d \Delta \tl{G}_\la\right\}~,\nn\\
A_\ka &=&\f{F_\phi}{16\pi^2}\left\{\tl{G}_{\ka}
-3d \Delta \tl{G}_S\right\}
\nn.\eeqa
with the deflection parameter $d=-2$. Here the discontinuities of Yukawa beta-function are given by
\beqa
\Delta G_{y_{ijk}}\equiv -\f{1}{8\pi^2}\Delta \tl{G}_{y_{ijk}}~,\nn
\eeqa
with
\beqa
\tl{G}_{\la}&=&4\la^2+2\ka^2+3y_t^2+3y_b^2+y_\tau^2-(3g_2^2+\f{3}{5}g_1^2)~,\\
\tl{G}_{\ka}&=&6\la^2+6\ka^2~,\nn\\
\tl{G}_{y_t}&=&\la^2+6y_t^2+y_b^2-(\f{16}{3}g_3^2+3g_2^2+\f{13}{15}g_1^2)~,\nn\\
\Delta \tl{G}_{y_t}&=&\la^2_{Q,3}\cos^2\psi_2+\la^2_{U,3}\cos^2\psi_2+\la^2\cot^2\theta+ 9y_t^2\tan^2\psi_2+3y_t^2\tan^4\psi_2+y_b^2\tan^2\psi_1,~\nn\\
\tl{G}_{y_b}&=&\la^2+y_t^2+6y_b^2+y_\tau^2-(\f{16}{3}g_3^2+3g_2^2+\f{7}{15}g_1^2)~,\nn\\
\Delta \tl{G}_{y_b}&=&\la^2_{Q,3}\cos^2\psi_2+\la^2_{D,3}\cos^2\psi_1+\la^2\cot^2\theta+ y_t^2\tan^2\psi_2+5y_b^2\tan^2\psi_2+4y_b^2\tan^2\psi_1~\nn\\
&+&3y_b^2\tan^2\psi_1\tan^2\psi_2+y_\tau^2\(\tan^2\psi_1+\tan^2\psi_2+\tan^2\psi_1\tan^2\psi_2\)~,\nn\\
\tl{G}_{y_\tau}&=&\la^2+3y_b^2+4y_\tau^2-(3g_2^2+\f{9}{5}g_1^2)~,\nn\\
\Delta\tl{G}_{y_\tau}&=&\la^2_{L,3}\cos^2\psi_1+\la^2_{E,3}\cos^2\psi_2+\la^2\cot^2\theta+3y_\tau^2\tan^2\psi_1+2y_\tau^2\tan^2\psi_2\nn\\
&+&y_\tau^2\tan^2\psi_1\tan^2\psi_2+3y_b^2\(\tan^2\psi_1+\tan^2\psi_2+\tan^2\psi_1\tan^2\psi_2\)~,\nn\\
\Delta \tl{G}_\la&=&\sin^2\theta\sum\limits_{a=1,2,3}\(6\la_{Q,a}^2+3\la_{U,a}^2+\la_{E,a}^2+3\la_{D,a}^2+2\la_{L,a}^2\)\nn\\
&+& 2\la^2\cot^2\theta+4\ka^2\cot^2\theta+2\ka^2\cot^4\theta+6y_t^2\tan^2\psi_2+3y_t^2\tan^4\psi_2~ \nn\\
&+&\(3y_b^2+y_\tau^2\)\(\tan^2\psi_1+\tan^2\psi_2+\tan^2\psi_1\tan^2\psi_2\),\nn\\
\Delta \tl{G}_S&=&\sin^2\theta\sum\limits_{a=1,2,3}\(6\la_{Q,a}^2+3\la_{U,a}^2+\la_{E,a}^2+3\la_{D,a}^2+2\la_{L,a}^2\)\nn\\
 &+&4\ka^2\cot^2\theta+2\ka^2\cot^4\theta~.
\eeqa
Within the expressions, we have used the relation $\tl{\la}\sin\theta=\la, \tl{\ka}\sin^3\theta=\ka$.

The soft SUSY breaking parameters
\beqa
m_{soft}^2=\delta_{d}+\delta_{I}+\delta_G~,
\eeqa
are given separately by
\bit
\item Pure deflected anomaly mediation contribution without new yuakwa couplings

\beqa
\delta^d_{{H}_u}~~&=&\f{F_\phi^2}{16\pi^2}\[\f{3}{2}G_2\al^2_2+\f{3}{10}G_1\al^2_1\]
+\f{F_\phi^2}{(16\pi^2)^2}\[\la^2\tl{G}_\la+3y_t^2\tl{G}_{y_t}\]~,~\nn\\
\delta^d_{{H}_d}~~&=&\f{F_\phi^2}{16\pi^2}\[\f{3}{2}G_2\al^2_2+\f{3}{10}G_1\al^2_1\]
+\f{F_\phi^2}{(16\pi^2)^2}\[\la^2\tl{G}_\la+3y_b^2\tl{G}_{y_b}+y_\tau^2\tl{G}_{y_\tau}\]~,~\nn\\
\delta^d_{\tl{Q}_{L;1,2}}&=&\f{F_\phi^2}{16\pi^2}\[\f{8}{3} G_3 \al^2_3+\f{3}{2}G_2\al^2_2+\f{1}{30}G_1\al^2_1\]~,~\nn\\
\delta^d_{\tl{U}^c_{L;1,2}}&=&\f{F_\phi^2}{16\pi^2}\[\f{8}{3} G_3 \al^2_3+\f{8}{15}G_1\al^2_1\]~,~\nn\\
\delta^d_{\tl{D}^c_{L;1,2}}&=&\f{F_\phi^2}{16\pi^2}\[\f{8}{3} G_3 \al^2_3+\f{2}{15}G_1\al^2_1\]~,~\nn\\
\delta^d_{\tl{L}_{L;1,2}}&=&\f{F_\phi^2}{16\pi^2}\[\f{3}{2}G_2\al_2^2+\f{3}{10}G_1\al_1^2\]~,~\nn\\
\delta^d_{\tl{E}_{L;1,2}^c}&=&\f{F_\phi^2}{16\pi^2}\f{6}{5}G_1\al_1^2~,~
\eeqa
with
\beqa
G_i&=&-b_i~,~\nn\\
(b_1,b_2,b_3)&=&(\f{33}{5},1,-3)~.
\eeqa
The soft SUSY breaking sfermions masses for the third generation needs the inclusion of the Yukawa contributions
\beqa
\delta^d_{\tl{Q}_{L,3}}&=&\delta^d_{\tl{Q}_{L;1,2}}+F_\phi^2\f{1}{(16\pi^2)^2}\[y_t^2\tl{G}_{y_t}+y_b^2\tl{G}_{y_b}\]~,\nn\\
\delta^d_{\tl{U}^c_{L,3}}&=&\delta^d_{\tl{U}^c_{L;1,2}}+F_\phi^2\f{1}{(16\pi^2)^2}\[2y_t^2\tl{G}_{y_t}\]~,\nn\\
\delta^d_{\tl{D}^c_{L,3}}&=&\delta^d_{\tl{D}^c_{L;1,2}}+F_\phi^2\f{1}{(16\pi^2)^2}\[2y_b^2\tl{G}_{y_b}\]~,~\nn\\
\delta^d_{\tl{L}_{L,3}}&=&\delta^d_{\tl{L}_{L;1,2}}+F_\phi^2\f{1}{(16\pi^2)^2}\[y_\tau^2\tl{G}_{y_\tau}\]~,~\nn\\
\delta^d_{\tl{E}_{L,3}^c}&=&\delta^d_{\tl{E}_{L;1,2}^c}+F_\phi^2\f{1}{(16\pi^2)^2}\[2y_\tau^2\tl{G}_{y_\tau}\]~,~
\eeqa

The pure anomaly contribution to the singlet soft masses $m_S^2$:
\beqa
\delta^d_{S}&=&\f{F_\phi^2}{(16\pi^2)^2}\[2\la^2\tl{G}_{\la}+2\ka^2\tl{G}_{\ka}\]~.
\eeqa

\item  The sum of GMSB type contributions $\delta_I$ with the interference contributions $\delta_G$.
Because the second term of GMSB contributions will cancel the contributions from the gauge-anomaly interference terms with $d=-2$, we give only the first term of the gauge mediated contributions for the NMSSM superfields:
\small
\beqa
\delta_{\tl{Q}_{L,a}}&=&\f{d^2 F_\phi^2}{(16\pi^2)^2}\[y^2_{Q Q_{L,a}\tl{S}_1}\(G_{y_{Q Q_{L,a}\tl{S}_1}}^+\)
 +y^2_{Q Q_{L,a}\tl{S}_0}\(G_{y_{Q Q_{L,a}\tl{S}_0}}^+\)\right.\nn\\
 &&~~~~+\left.\delta_{a,3} y^2_{Q_{L,3} K_U H_u}\(G_{y_{Q_{L,3} K_U H_u}}^+\)+\delta_{a,3} y^2_{Q_{L,3} K_D H_d}\(G_{y_{Q_{L,3} K_D H_d}}^+\)\],~~\nn\\
\delta_{\tl{U}^c_{L,a}}&=&\f{d^2 F_\phi^2}{(16\pi^2)^2}\[y^2_{U U^c_{L,a}\tl{S}_1}\(G_{y_{U U^c_{L,a}\tl{S}_1}}^+\)
 +y^2_{U U^c_{L,a}\tl{S}_0}\(G_{y_{U U^c_{L,a}\tl{S}_1}}^+\)+\delta_{a,3} 2y^2_{\bar{K}_{Q} U_{L,3}^c H_u}\(G_{y_{\bar{K}_{Q} U_{L,3}^c H_u}}^+\)\],~~\nn\\
\delta_{\tl{E}_{L,a}^c}&=&\f{d^2 F_\phi^2}{(16\pi^2)^2}\[y^2_{E E^c_{L,a}\tl{S}_1}\(G_{y_{E E^c_{L,a}\tl{S}_1}}^+\)
 +y^2_{E E^c_{L,a}\tl{S}_0}\(G_{y_{E E^c_{L,a}\tl{S}_0}}^+\)+\delta_{a,3} 2y^2_{ {K}_{L} E_{L,3}^c H_d}\(G_{y_{ {K}_{L} E_{L,3}^c H_d}}^+\)\],~~\nn\\
\delta_{\tl{D}^c_{L,a}}&=&\f{d^2 F_\phi^2}{(16\pi^2)^2}\[y^2_{D D^c_{L,a}\tl{S}_1}\(G_{y_{D D^c_{L,a}\tl{S}_1}}^+\)
 +y^2_{D D^c_{L,a}\tl{S}_0}\(G_{y_{D D^c_{L,a}\tl{S}_0}}^+\)+\delta_{a,3} 2y^2_{ \bar{K}_{Q} D_{L,3}^c H_d}\(G_{y_{\bar{K}_{Q} D_{L,3}^c H_d}}^+\)\],~~\nn\\
\delta_{\tl{L}_{L,a}}&=&\f{d^2 F_\phi^2}{(16\pi^2)^2}\[y^2_{L L_{L,a}\tl{S}_1}\(G_{y_{L L_{L,a}\tl{S}_1}}^+\)
 +y^2_{L L_{L,a}\tl{S}_0}\(G_{y_{L L_{L,a}\tl{S}_0}}^+\)+\delta_{a,3} y^2_{ \bar{L}_{L,3} {K}_{E} H_d}\(G_{y_{\bar{L}_{L,3} {K}_{E} H_d}}^+\)\],\nn\\
\delta_{H_u}~&=&\f{d^2 F_\phi^2}{(16\pi^2)^2}\[y^2_{\tl{S}_0H_d H_u}\(G_{y_{\tl{S}_0H_d H_u}}^+\)+3y^2_{Q_{L,3} K_U H_u}\(G_{y_{Q_{L,3} K_U H_u}}^+\)+3y^2_{{K}_QU_{L,3}^c H_u}\(G_{y_{{K}_QU_{L,3}^c H_u}}^+\)\right.~\nn\\
&&~~~~~~~~~~+\left.3y^2_{{K}_Q K_U H_u}\(G_{y_{{K}_Q K_U H_u}}^+\)\]~,\nn\\
\delta_{H_d}~&=&\f{d^2 F_\phi^2}{(16\pi^2)^2}\[y^2_{\tl{S}_0H_d H_u}\(G_{y_{\tl{S}_0H_d H_u}}^+\)+3y^2_{Q_{L,3} K_D H_d}\(G_{y_{Q_{L,3} K_D H_d}}^+\)+3y^2_{{K}_Q D_{L,3}^c H_d}\(G_{y_{{K}_Q D_{L,3}^c H_d}}^+\)\right.\nn\\
&&~~~~~~~~~+3y^2_{{K}_Q K_D H_d}\(G_{y_{{K}_Q K_D H_d}}^+\)+y^2_{\bar{K}_L E_{L,3}^c H_d}\(G_{y_{\bar{K}_L E_{L,3}^c H_d}}^+\)+y^2_{\bar{L}_{L,3} {K}_E H_d}\(G_{y_{\bar{L}_{L,3} {K}_E H_d}}^+\)\nn\\
&&~~~~~~~~~~\left.+y^2_{\bar{K}_{L} {K}_E H_d}\(G_{y_{\bar{K}_{L} {K}_E H_d}}^+\)\],\nn\\
\delta_{S}&=&\f{d^2 F_\phi^2}{(16\pi^2)^2}\left\{\[\sum\limits_{a=1,2,3}\(
6y_{Q Q_{L,a}\tl{S}_1}^2\(G_{y_{Q Q_{L,a}\tl{S}_1}}^+\)+3 y_{U U^c_{L,a}\tl{S}_1}^2\(G_{y_{U U^c_{L,a}\tl{S}_1}}^+\)
\right.\right.\right.\nn\\
&&\left.+y_{E E^c_{L,a}\tl{S}_1}^2\(G_{y_{E E^c_{L,a}\tl{S}_1}}^+\)
+3y^2_{D D^c_{L,a}\tl{S}_1}\(G_{y_{D D^c_{L,a}\tl{S}_1}}^+\)+2y^2_{L L_{L,a}\tl{S}_1}\(G_{y_{L L_{L,a}\tl{S}_1}}^+\)\]\nn\\
&&+6y_{Q K_Q\tl{S}_1}^2\(G_{y_{Q K_Q \tl{S}_1}}^+\)+  3 y_{U K_U \tl{S}_1}^2\(G_{y_{U K_U \tl{S}_1}}^+\)+y_{E K_E\tl{S}_1}^2\(G_{y_{E K_E\tl{S}_1}}^+\)\nn\\
&&+3y^2_{D K_D \tl{S}_1}\(G_{y_{D K_D \tl{S}_1}}^+\)
+2y^2_{L K_L\tl{S}_1}\(G_{y_{L K_L \tl{S}_1}}^+\)+4y_{\tl{S}_0\tl{S}_1\tl{S}_1}^2\(G_{y_{\tl{S}_0\tl{S}_1\tl{S}_1}}^+\)+2y_{\tl{S}_0\tl{S}_0\tl{S}_1}^2\(G_{y_{\tl{S}_0\tl{S}_0\tl{S}_1}}^+\)\left.\f{}{}\right\},\nn
\eeqa
\normalsize
with $d=-2$ and $\delta_{a,3}$ the Kronecker delta. We identify the $S$ field in NMSSM with $\tl{S}_1$ in our scenario.

 The anomalous dimensions upon the messenger threshold are given as
\beqa
G_{y_{Q Q_{L,b}\tl{S}_1}}^+&=&
 \sin^2\theta\sum\limits_{a=1,2,3}\[6\la_{Q,a}^2+3\la_{U,a}^2+\la_{E,a}^2+3(\la_{D,a})^2+2(\la_{L,a})^2\]\nn\\
               &&~~~~+\sum\limits_{a=1,2,3}(\la_{Q,a}^2)+2\ka^2\cot^4\theta+4\ka^2\cot^2\theta+ 2\la^2+2\ka^2~\nn\\
               &&~~~~+\la^2_{Q,b}(\delta_{b,1}+\delta_{b,2})+\delta_{b,3}(\la^2_{Q,3}\cos^2\psi_2+ y_t^2\tan^2\psi_2+y_b^2\tan^2\psi_1
               +y_t^2+y_b^2)\nn\\
               &&~~~~- \f{16}{3}g_3^2-{3}g_2^2-\f{1}{15}g_1^2~,\\
G_{y_{Q K_Q\tl{S}_1}}^+&=&
 \sin^2\theta\sum\limits_{a=1,2,3}\[6\la_{Q,a}^2+3\la_{U,a}^2+\la_{E,a}^2+3(\la_{D,a})^2+2(\la_{L,a})^2\]\nn\\
               &&~~~~+\sum\limits_{a=1,2,3}(\la_{Q,a}^2)+4\ka^2\cot^4\theta+2\ka^2\cot^2\theta+2\la^2\cot^2\theta+2\ka^2\cot^6\theta~\nn\\
               &&~~~~+\la^2_{Q,3}\sin^2\psi_2+ y_t^2\tan^4\psi_2+y_b^2\tan^2\psi_1\tan^2\psi_2+(y_t^2+y_b^2)\tan^2\psi_2\nn\\
               &&~~~~-\f{16}{3}g_3^2-{3}g_2^2-\f{1}{15}g_1^2~,\\
G_{y_{Q Q_{L,b}\tl{S}_0}}^+&=&
 \cos^2\theta\sum\limits_{a=1,2,3}\[6\la_{Q,a}^2+3\la_{U,a}^2+\la_{E,a}^2+3(\la_{D,a})^2+2(\la_{L,a})^2\]\nn\\
               &&~~~~+\sum\limits_{a=1,2,3}(\la_{Q,a}^2)+4\ka^2\cot^4\theta+2\ka^2\cot^2\theta+2\la^2\cot^2\theta+2\ka^2\cot^6\theta~\nn\\
               &&~~~~+\la^2_{Q,b}(\delta_{b,1}+\delta_{b,2})+\delta_{b,3}(\la^2_{Q,3}\cos^2\psi_2+ y_t^2\tan^2\psi_2+y_b^2\tan^2\psi_1
               +y_t^2+y_b^2)\nn\\
               &&~~~~- \f{16}{3}g_3^2-{3}g_2^2-\f{1}{15}g_1^2~,\\
\hline\nn\\
G_{y_{U U^c_{L,b}\tl{S}_1}}^+&=&
\sin^2\theta\sum\limits_{a=1,2,3}\[6\la_{Q,a}^2+3\la_{U,a}^2+\la_{E,a}^2+3(\la_{D,a})^2+2(\la_{L,a})^2\]\nn\\
               &&~~~~+\sum\limits_{a=1,2,3}(\la_{U,a}^2)+2\ka^2\cot^4\theta+4\ka^2\cot^2\theta+ 2\la^2+2\ka^2~\nn\\
               &&~~~~+\la^2_{U,b}(\delta_{b,1}+\delta_{b,2})+\delta_{b,3}(\la^2_{U,3}\cos^2\psi_2+2y_t^2\tan^2\psi_2+2y_t^2)\nn\\
               &&~~~~- \f{16}{3}g_3^2-\f{16}{15}g_1^2~,\\
G_{y_{U K_U\tl{S}_1}}^+&=&
\sin^2\theta\sum\limits_{a=1,2,3}\[6\la_{Q,a}^2+3\la_{U,a}^2+\la_{E,a}^2+3(\la_{D,a})^2+2(\la_{L,a})^2\]\nn\\
               &&~~~~+\sum\limits_{a=1,2,3}(\la_{U,a}^2)+2\ka^2\cot^4\theta+4\ka^2\cot^2\theta+ 2\la^2+2\ka^2~\nn\\
               &&~~~~+\la^2_{U,3}\sin^2\psi_2+2y_t^2\tan^2\psi_2+2y_t^2\tan^4\psi_2\nn\\
               &&~~~~- \f{16}{3}g_3^2-\f{16}{15}g_1^2~,\\
 G_{y_{U U^c_{L,b}\tl{S}_0}}^+&=&\cos^2\theta\sum\limits_{a=1,2,3}\[6\la_{Q,a}^2+3\la_{U,a}^2+\la_{E,a}^2+3(\la_{D,a})^2+2(\la_{L,a})^2\]\nn\\
               &&~~~~+\sum\limits_{a=1,2,3}(\la_{U,a}^2)+4\ka^2\cot^4\theta+2\ka^2\cot^2\theta+ 2\la^2\cot^2\theta+2\ka^2\cot^6\theta~\nn\\
               &&~~~~+\la^2_{U,b}(\delta_{b,1}+\delta_{b,2})+\delta_{b,3}(\la^2_{U,3}\cos^2\psi_2+2y_t^2\tan^2\psi_2+2y_t^2)\nn\\
               &&~~~~- \f{16}{3}g_3^2-\f{16}{15}g_1^2~,\\
\hline\nn\\
G_{y_{D D^c_{L,b}\tl{S}_1}}^+&=&
 \sin^2\theta\sum\limits_{a=1,2,3}\[6\la_{Q,a}^2+3\la_{U,a}^2+\la_{E,a}^2+3(\la_{D,a})^2+2(\la_{L,a})^2\]\nn\\
               &&~~~~+\sum\limits_{a=1,2,3}(\la_{D,a}^2)+2\ka^2\cot^4\theta+4\ka^2\cot^2\theta+ 2\la^2+2\ka^2~\nn\\
               &&~~~~+\la^2_{D,b}(\delta_{b,1}+\delta_{b,2})+\delta_{b,3}(\la^2_{D,3}\cos^2\psi_1+2y_b^2\tan^2\psi_2+2y_b^2)\nn\\
               &&~~~~-\f{16}{3}g_3^2-\f{4}{15}g_1^2~,\\
G_{y_{D K_D \tl{S}_1}}^+&=&
 \sin^2\theta\sum\limits_{a=1,2,3}\[6\la_{Q,a}^2+3\la_{U,a}^2+\la_{E,a}^2+3(\la_{D,a})^2+2(\la_{L,a})^2\]\nn\\
               &&~~~~+\sum\limits_{a=1,2,3}(\la_{D,a}^2)+2\ka^2\cot^4\theta+4\ka^2\cot^2\theta+ 2\la^2+2\ka^2~\nn\\
               &&~~~~+\la^2_{D,3}\sin^2\psi_1+2y_b^2\tan^2\psi_1+2y_b^2\tan^2\psi_1\tan^2\psi_2\nn\\
               &&~~~~-\f{16}{3}g_3^2-\f{4}{15}g_1^2~,\\
G_{y_{D D^c_{L,b}\tl{S}_0}}^+&=&\cos^2\theta\sum\limits_{a=1,2,3}\[6\la_{Q,a}^2+3\la_{U,a}^2+\la_{E,a}^2+3(\la_{D,a})^2+2(\la_{L,a})^2\]\nn\\
               &&~~~~+\sum\limits_{a=1,2,3}(\la_{D,a}^2)+4\ka^2\cot^4\theta+2\ka^2\cot^2\theta+ 2\la^2\cot^2\theta+2\ka^2\cot^6\theta~\nn\\
               &&~~~~+\la^2_{D,b}(\delta_{b,1}+\delta_{b,2})+\delta_{b,3}(\la^2_{D,3}\cos^2\psi_1+2y_b^2\tan^2\psi_2+2y_b^2)\nn\\
               &&~~~~-\f{16}{3}g_3^2-\f{4}{15}g_1^2~,\\
\hline\nn\\
G_{y_{L L_{L,b}\tl{S}_1}}^+&=&
\sin^2\theta\sum\limits_{a=1,2,3}\[6\la_{Q,a}^2+3\la_{U,a}^2+\la_{E,a}^2+3(\la_{D,a})^2+2(\la_{L,a})^2\]\nn\\
               &&~~~~+\sum\limits_{a=1,2,3}(\la_{L,a}^2)+2\ka^2\cot^4\theta+4\ka^2\cot^2\theta+ 2\la^2+2\ka^2~\nn\\
               &&~~~~+\la^2_{L,b}(\delta_{b,1}+\delta_{b,2})+\delta_{b,3}(\la^2_{L,3}\cos^2\psi_1+y_\tau^2\tan^2\psi_2+y_\tau^2)\nn\\
               &&~~~~-{3}g_2^2-\f{3}{5}g_1^2~,\\
G_{y_{L K_L\tl{S}_1}}^+&=&
\sin^2\theta\sum\limits_{a=1,2,3}\[6\la_{Q,a}^2+3\la_{U,a}^2+\la_{E,a}^2+3(\la_{D,a})^2+2(\la_{L,a})^2\]\nn\\
               &&~~~~+\sum\limits_{a=1,2,3}(\la_{L,a}^2)+2\ka^2\cot^4\theta+4\ka^2\cot^2\theta+ 2\la^2+2\ka^2~\nn\\
               &&~~~~+\la^2_{L,3}\sin^2\psi_1+y_\tau^2\tan^2\psi_1+y_\tau^2\tan^2\psi_1\tan^2\psi_2\nn\\
               &&~~~~-{3}g_2^2-\f{3}{5}g_1^2~,\\
G_{y_{L L_{L,b}\tl{S}_0}}^+&=&\cos^2\theta\sum\limits_{a=1,2,3}\[6\la_{Q,a}^2+3\la_{U,a}^2+\la_{E,a}^2+3(\la_{D,a})^2+2(\la_{L,a})^2\]\nn\\
               &&~~~~+\sum\limits_{a=1,2,3}(\la_{L,a}^2)+4\ka^2\cot^4\theta+2\ka^2\cot^2\theta+ 2\la^2\cot^2\theta+2\ka^2\cot^6\theta~\nn\\
              &&~~~~+\la^2_{L,b}(\delta_{b,1}+\delta_{b,2})+\delta_{b,3}(\la^2_{L,3}\cos^2\psi_1+y_\tau^2\tan^2\psi_2+y_\tau^2)\nn\\
               &&~~~~-{3}g_2^2-\f{3}{5}g_1^2~,\\
\hline\nn\\
G_{y_{E E^c_{L,b}\tl{S}_1}}^+&=&
 \sin^2\theta\sum\limits_{a=1,2,3}\[6\la_{Q,a}^2+3\la_{U,a}^2+\la_{E,a}^2+3(\la_{D,a})^2+2(\la_{L,a})^2\]\nn\\
               &&~~~~+\sum\limits_{a=1,2,3}(\la_{E,a}^2)+2\ka^2\cot^4\theta+4\ka^2\cot^2\theta+ 2\la^2+2\ka^2~\nn\\
               &&~~~~+\la^2_{E,b}(\delta_{b,1}+\delta_{b,2})+\delta_{b,3}(\la^2_{E,3}\cos^2\psi_2+2y_\tau^2\tan^2\psi_1+2y_\tau^2)-\f{12}{5}g_1^2~,\\
G_{y_{E K_E\tl{S}_1}}^+&=&
 \sin^2\theta\sum\limits_{a=1,2,3}\[6\la_{Q,a}^2+3\la_{U,a}^2+\la_{E,a}^2+3(\la_{D,a})^2+2(\la_{L,a})^2\]\nn\\
               &&~~~~+\sum\limits_{a=1,2,3}(\la_{E,a}^2)+2\ka^2\cot^4\theta+4\ka^2\cot^2\theta+ 2\la^2+2\ka^2~\nn\\
               &&~~~~+\la^2_{E,3}\sin^2\psi_2+2y_\tau^2\tan^2\psi_2+2y_\tau^2\tan^2\psi_1\tan^2\psi_2-\f{12}{5}g_1^2~,\\
G_{y_{E E^c_{L,b}\tl{S}_0}}^+&=&\cos^2\theta\sum\limits_{a=1,2,3}\[6\la_{Q,a}^2+3\la_{U,a}^2+\la_{E,a}^2+3(\la_{D,a})^2+2(\la_{L,a})^2\]\nn\\
               &&~~~~+\sum\limits_{a=1,2,3}(\la_{E,a}^2)+4\ka^2\cot^4\theta+2\ka^2\cot^2\theta+ 2\la^2\cot^2\theta+2\ka^2\cot^6\theta~\nn\\
               &&~~~~+\la^2_{E,b}(\delta_{b,1}+\delta_{b,2})+\delta_{b,3}(\la^2_{E,3}\cos^2\psi_2+2y_\tau^2\tan^2\psi_1+2y_\tau^2)-\f{12}{5}g_1^2~,\\
\hline\nn\\
G_{y_{\tl{S}_0H_d H_u}}^+&=&
\cos^2\theta\sum\limits_{a=1,2,3}\[6\la_{Q,a}^2+3\la_{U,a}^2+\la_{E,a}^2+3(\la_{D,a})^2+2(\la_{L,a})^2\]\nn\\
               &&~~+4\ka^2\cot^4\theta+2\ka^2\cot^2\theta+2\ka^2\cot^6\theta+4\la^2\cot^2\theta+ 2\la^2+6y_t^2\tan^2\psi_2~\nn\\
               &&~~+3y_t^2\tan^4\psi_2+\(3y_b^2+y_\tau^2\)\(\tan^2\psi_1+\tan^2\psi_2+\tan^2\psi_1\tan^2\psi_2\)\nn\\
               &&~~~~~~~~+3y_t^2+3y_b^2+y_\tau^2-{3}g_2^2-\f{3}{5}g_1^2~,\nn\\
G_{y_{\tl{S}_0\tl{S}_1\tl{S}_1}}^+&=&
\cos^2\theta\sum\limits_{a=1,2,3}\[6\la_{Q,a}^2+3\la_{U,a}^2+\la_{E,a}^2+3(\la_{D,a})^2+2(\la_{L,a})^2\]\nn\\
               &&~~~~~~~+4\ka^2\cot^4\theta+2\ka^2\cot^2\theta+2\ka^2\cot^6\theta+2\la^2\cot^2\theta~,\nn\\
               &+&2\sin^2\theta\sum\limits_{a=1,2,3}\[6\la_{Q,a}^2+3\la_{U,a}^2+\la_{E,a}^2+3(\la_{D,a})^2+2(\la_{L,a})^2\]\nn\\
               &&~~~~~~~+2\[2\ka^2\cot^4\theta+4\ka^2\cot^2\theta+ 2\ka^2+2\la^2\]~,\nn\\
G_{y_{\tl{S}_0\tl{S}_0\tl{S}_1}}^+&=&
2\cos^2\theta\sum\limits_{a=1,2,3}\[6\la_{Q,a}^2+3\la_{U,a}^2+\la_{E,a}^2+3(\la_{D,a})^2+2(\la_{L,a})^2\]\nn\\
               &&~~~~~~~+2\[4\ka^2\cot^4\theta+2\ka^2\cot^2\theta+2\ka^2\cot^6\theta+2\la^2\cot^2\theta\]~,\nn\\
               &+&\sin^2\theta\sum\limits_{a=1,2,3}\[6\la_{Q,a}^2+3\la_{U,a}^2+\la_{E,a}^2+3(\la_{D,a})^2+2(\la_{L,a})^2\]\nn\\
               &&~~~~~~~+ \[2\ka^2\cot^4\theta+4\ka^2\cot^2\theta+ 2\ka^2+2\la^2\]~,\\
\hline\nn\\
G_{y_{Q_{L,3} K_U H_u}}^+&=&\la^2_{Q,3}\cos^2\psi_2+\la_{U,3}^2\sin^2\psi_2+9y_t^2\tan^2\psi_2+5y_t^2\tan^4\psi_2+y_b^2\tan^2\psi_1\nn\\
&&+\la^2\cot^2\theta+4y_t^2+y_b^2-\f{16}{3}g_3^2-3g_2^2-\f{13}{15}g_1^2~,\nn\\
%
G_{y_{\bar{K}_{Q} U_{L,3}^c H_u}}^+&=&\la^2_{Q,3}\sin^2\psi_2+\la_{U,3}^2\cos^2\psi_2+(9y_t^2+y_b^2)\tan^2\psi_2+ 4y_t^2\tan^4\psi_2+y_b^2\tan^2\psi_1\tan^2\psi_2\nn\\
&&+\la^2\cot^2\theta+5y_t^2-\f{16}{3}g_3^2-3g_2^2-\f{13}{15}g_1^2~,\nn\\
G_{y_{\bar{K}_{Q} K_U H_u}}^+&=&\la^2_{Q,3}\sin^2\psi_2+\la_{U,3}^2\sin^2\psi_2+(9y_t^2+y_b^2)\tan^2\psi_2+ 6y_t^2\tan^4\psi_2+y_b^2\tan^2\psi_1\tan^2\psi_2\nn\\
&&+\la^2\cot^2\theta-\f{16}{3}g_3^2-3g_2^2-\f{13}{15}g_1^2~,\\
%
\hline\nn\\
G_{y_{Q_{L,3} K_D H_d}}^+&=&\la^2_{Q,3}\cos^2\psi_2+\la_{D,3}^2\sin^2\psi_1+y_t^2\tan^2\psi_2+3y_b^2\tan^2\psi_1+2y_b^2\tan^2\psi_1\tan^2\psi_2\nn\\
&&+(3y_b^2+y_\tau^2)\(\tan^2\psi_1+\tan^2\psi_2+\tan^2\psi_1\tan^2\psi_2\)\nn\\
&&+\la^2\cot^2\theta+y_t^2+4y_b^2+y_\tau^2-\f{16}{3}g_3^2-3g_2^2-\f{7}{15}g_1^2~,\nn\\
G_{y_{\bar{K}_{Q} D_{L,3}^c H_d}}^+&=&\la^2_{Q,3}\sin^2\psi_2+\la_{D,3}^2\cos^2\psi_1+y_t^2\tan^2\psi_2+y_t^2\tan^4\psi_2+3y_b^2\tan^2\psi_2+y_b^2\tan^2\psi_1\tan^2\psi_2\nn\\
&&+(3y_b^2+y_\tau^2)\(\tan^2\psi_1+\tan^2\psi_2+\tan^2\psi_1\tan^2\psi_2\)\nn\\
&&+\la^2\cot^2\theta+5y_b^2+y_\tau^2-\f{16}{3}g_3^2-3g_2^2-\f{7}{15}g_1^2~,\nn\\
G_{y_{K_Q K_D H_d}}^+&=&\la^2_{Q,3}\sin^2\psi_2+\la_{D,3}^2\sin^2\psi_1+y_t^2\tan^2\psi_2+y_t^2\tan^4\psi_2+4y_b^2\tan^2\psi_2+6y_b^2\tan^2\psi_1\tan^2\psi_2\nn\\
&&+5y_b^2\tan^2\psi_1+y_\tau^2\(\tan^2\psi_1+\tan^2\psi_2+\tan^2\psi_1\tan^2\psi_2\)\nn\\
&&+\la^2\cot^2\theta-\f{16}{3}g_3^2-3g_2^2-\f{7}{15}g_1^2~,\\
\hline\nn\\
G_{y_{\bar{L}_{L,3} {K}_{E} H_d}}^+&=&\la_{L,3}^2\cos^2\psi_1+\la^2_{E,3}\sin^2\psi_1+4y_\tau^2\tan^2\psi_2+3y_\tau^2\tan^2\psi_1\tan^2\psi_2+y_\tau^2\tan^2\psi_1\nn\\
&&+3y_b^2\(\tan^2\psi_1+\tan^2\psi_2+\tan^2\psi_1\tan^2\psi_2\)\nn\\
&&+\la^2\cot^2\theta+3y_b^2+2y_\tau^2-3g_2^2-\f{9}{5}g_1^2~,\nn\\
G_{y_{{K}_{L} E_{L,3}^c H_d}}^+&=&\la^2_{L,3}\sin^2\psi_1+\la_{E,3}^2\cos^2\psi_2+y_\tau^2\tan^2\psi_2+2y_\tau^2\tan^2\psi_1\tan^2\psi_2+4y_\tau^2\tan^2\psi_1\nn\\
&&+3y_b^2\(\tan^2\psi_1+\tan^2\psi_2+\tan^2\psi_1\tan^2\psi_2\)\nn\\
&&+\la^2\cot^2\theta+3y_b^2+3y_\tau^2-3g_2^2-\f{9}{5}g_1^2~,\nn\\
G_{y_{\bar{K}_L {K}_{E} H_d}}^+&=&\la_{L,3}^2\sin^2\psi_1+\la^2_{E,3}\sin^2\psi_1+3y_\tau^2\tan^2\psi_2+4y_\tau^2\tan^2\psi_1\tan^2\psi_2+2y_\tau^2\tan^2\psi_1\nn\\
&&+3y_b^2\(\tan^2\psi_1+\tan^2\psi_2+\tan^2\psi_1\tan^2\psi_2\)\nn\\
&&+\la^2\cot^2\theta-3g_2^2-\f{9}{5}g_1^2~,\nn\\
\hline\nn\\
\eeqa
\eit

\end{document}